\documentclass[sigconf]{acmart}
\usepackage{booktabs,multirow}
\usepackage{enumitem}
\usepackage{float}
\usepackage{pifont}
\usepackage[utf8]{inputenc}
\newcommand{\cmark}{\ding{51}}
\AtBeginDocument{%
  }

\setcopyright{acmlicensed}
\copyrightyear{2018}
\acmYear{2018}
\acmDOI{XXXXXXX.XXXXXXX}
\acmConference[Conference acronym 'XX]{Make sure to enter the correct
  conference title from your rights confirmation email}{June 03--05,
  2018}{Woodstock, NY}
\acmISBN{978-1-4503-XXXX-X/2018/06}

\begin{document}

\title{Q-BERT4Rec: Quantized Semantic-ID Representation Learning for Multimodal Recommendation}

\author{Haofeng Huang\textsuperscript{1}}
\affiliation{%
  \institution{University of Shanghai for Science and Technology}
  \city{Shanghai}
  \country{China}
}
\email{2235060414@st.usst.edu.cn}

\author{Ling Gai\textsuperscript{2,*}}
\affiliation{%
  \institution{University of Shanghai for Science and Technology}
  \city{Shanghai}
  \country{China}
}
\email{lgai@usst.edu.cn}

\thanks{\textsuperscript{*} Corresponding author.}


\begin{abstract}
Sequential recommendation plays a critical role in modern online platforms such as e-commerce, advertising, and content streaming, where accurately predicting users’ next interactions is essential for personalization. Recent Transformer-based methods like BERT4Rec have shown strong modeling capability, yet they still rely on discrete item IDs that lack semantic meaning and ignore rich multimodal information (e.g., text and image). This leads to weak generalization and limited interpretability. To address these challenges, we propose Q-Bert4Rec, a multimodal sequential recommendation framework that unifies semantic representation and quantized modeling. Specifically, Q-Bert4Rec consists of three stages: (1) cross-modal semantic injection, which enriches randomly initialized ID embeddings through a dynamic transformer that fuses textual, visual, and structural features; (2) semantic quantization, which discretizes fused representations into meaningful tokens via residual vector quantization; and (3) multi-mask pretraining and fine-tuning, which leverage diverse masking strategies—span, tail, and multi-region—to improve sequential understanding. We validate our model on public Amazon benchmarks demonstrate that Q-Bert4Rec significantly outperforms many strong existing methods, confirming the effectiveness of semantic tokenization for multimodal sequential recommendation. Our source code will be publicly available on Github after publishing.
\end{abstract}



\maketitle
\begin{figure}[t]
    \centering
    \includegraphics[width=0.9\linewidth,trim=60 165 290 220,clip]{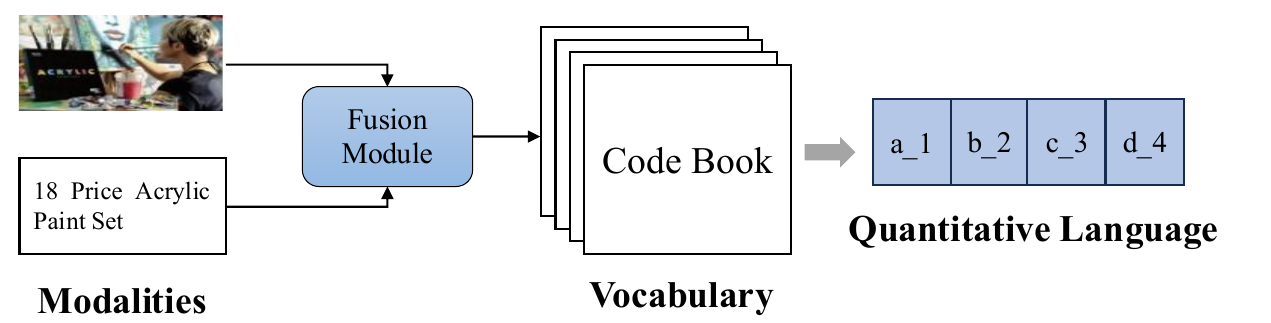}
    \caption{
        Overall framework of the proposed Semantic-ID Quantization. The model fuses multi-modal inputs through a fusion module and maps them into a shared quantized vocabulary space, forming discrete token sequences that serve as a compact and interpretable quantitative language
    }
    \label{fig:layers}
\end{figure}
\section{INTRODUCTION}
Recommend systems have become indispensable across e-commerce, digital advertising, and content streaming, where they surface relevant items from massive catalogs and drive key business metrics such as CTR, retention, and revenue\cite{zhang2019deep}. Among various paradigms, sequential recommendation focuses on modeling users’ historical interactions to predict their next actions, thereby capturing evolving preferences and short-term intents. Early neural methods such as GRU4Rec\cite{jannach2017recurrent} leveraged recurrent architectures to encode temporal dynamics, while self-attention models like SASRec\cite{kang2018self} improved long-range dependency modeling and parallelism. Building on the Transformer, BERT4Rec\cite{sun2019bert4rec} introduced bidirectional contextualization for sequences and set a strong foundation for many subsequent advances. These developments establish the Transformer as a de facto backbone for sequence modeling in recommendation.

To enhance representation quality, recent work focuses on exploring multimodal signals—text (titles, descriptions), images, and structured attributes—to enrich item semantics beyond bare IDs (e.g., VBPR, MMGCN, DualGNN)\cite{he2016vbpr, wei2019mmgcn, wang2021dualgnn}. In parallel, generative and prompt-based recommenders reframe recommendation as a language or sequence generation problem\cite{deldjoo2024review}, unifying diverse tasks through instruction tuning or tokenization (e.g., P5, M6-Rec, TIGER, VIP5)\cite{geng2022recommendation, cui2022m6, rajput2023recommender, geng2023vip5}. More recent work such as MQL4GRec\cite{zhai2025multimodal} and DiffMM\cite{jiang2024diffmm}further explored quantized or generative modeling to bridge continuous multimodal features and discrete semantics. MQL4GRec employs Residual Vector Quantization (RQ-VAE)~\cite{zhai2025multimodal} to convert multimodal representations into discrete semantic tokens. However, it performs quantization independently for each modality, which results in inconsistent codebook distributions across modalities. This misalignment weakens the shared semantic space and hinders the integration of multimodal information in sequential modeling. DiffMM, while leveraging diffusion-based reconstruction\cite{ho2020denoising} for generation, primarily focuses on denoising processes rather than fine-grained multimodal fusion. These examples reveal a fundamental challenge: existing models either perform static multimodal fusion without adaptivity, or decouple quantization from sequence modeling, thereby limiting both interpretability and generalization.

The development of Residual Vector Quantization (RQ-VAE) techniques\cite{oord2017vqvae, zeghidour2021soundstream} has further inspired us to address these limitations. Similarly to how pretraining–finetuning\cite{devlin2019bert} and prompt-tuning\cite{brown2020gpt3}  revolutionized natural language generation by introducing a shared vocabulary with transferable semantics, we envision a comparable paradigm shift for recommendation. Instead of representing items by arbitrary, meaningless IDs, we propose to replace them with semantic IDs—compact sequences of quantized semantic tokens that encapsulate multimodal meaning. These semantic IDs act as both identifiers and language units, bridging the gap between discrete recommendation tokens and rich multimodal semantics. In this view, each token from the quantized codebook serves as a semantic atom, much like words in a vocabulary that encode transferable priors across domains. Consequently, recommendation can be reformulated as a language of items, where interactions are modeled as sentences composed of semantic tokens. This new representation not only enables cross-domain knowledge transfer and interpretability, but also maintains efficiency and scalability by being more compact than original modalities such as text or images. As illustrated in Figure 1, the proposed semantic-ID paradigm provides a unified, token-level bridge connecting multimodal content, quantized representations, and sequential reasoning.

To this paper, we propose a novel Quantized BERT-style Multimodal Sequential Recommendation framework, named Q-Bert4Rec. Specifically, our method is motivated by the observation that traditional sequential recommenders represent items as discrete and meaningless IDs, which lack semantic understanding and hinder generalization across domains. To bridge this gap, Q-Bert4Rec introduces a unified three-stage framework that learns semantic IDs composed of quantized multimodal tokens, enabling both semantic richness and discrete modeling. In the first stage, we perform Cross-modal Semantic Injection, where a dynamic Transformer fuses textual, visual, and structural features into randomly initialized ID embeddings. A learnable gating mechanism adaptively controls the fusion depth, ensuring content-aware semantic enrichment for different items.In the second stage, we introduce Semantic Quantization using a Residual Vector Quantization module (RQ-VAE)\cite{zeghidour2021soundstream}. The fused representations are discretized into compact, interpretable semantic tokens, which serve as the new identifiers—semantic IDs—to replace original item IDs. These quantized tokens form a unified vocabulary that encodes multimodal meaning in a transferable and efficient way. In the third stage, we design a Multi-mask Pretraining and Fine-tuning strategy to fully exploit sequential dependencies. By combining span, tail, and multi-region masking, the model captures both local continuity and long-range correlations in user behavior sequences. This stage enhances the model’s ability to reason over diverse sequential patterns and improves its adaptability to downstream recommendation tasks. We evaluate Q-Bert4Rec on multiple public Amazon benchmarks, and the results show significant improvements compared to strong baselines. The findings demonstrate that unifying semantic fusion and quantized modeling effectively bridges the gap between continuous multimodal representations and discrete recommendation reasoning.

\noindent\textbf{Our main contributions are summarized as follows:}
\begin{itemize}[topsep=0pt, itemsep=1pt, parsep=0pt, leftmargin=*]
    \item We introduce Q-Bert4Rec, a novel quantized BERT-style multimodal sequential recommendation framework that unifies semantic representation learning and discrete token modeling.
    \item We propose a three-stage architecture consisting of a dynamic cross-modal semantic injection module for adaptive multimodal fusion;a residual vector quantization mechanism to transform fused embeddings into interpretable \emph{semantic IDs}; and a multi-mask pretraining strategy that enhances temporal reasoning and robustness.
    \item We conduct extensive experiments and analyses on three public datasets, and the results validate the effectiveness of our proposed method.
\end{itemize}

\begin{figure*}[t]
    \centering
    \includegraphics[width=0.9\linewidth, trim = 0 5 0 0, clip]{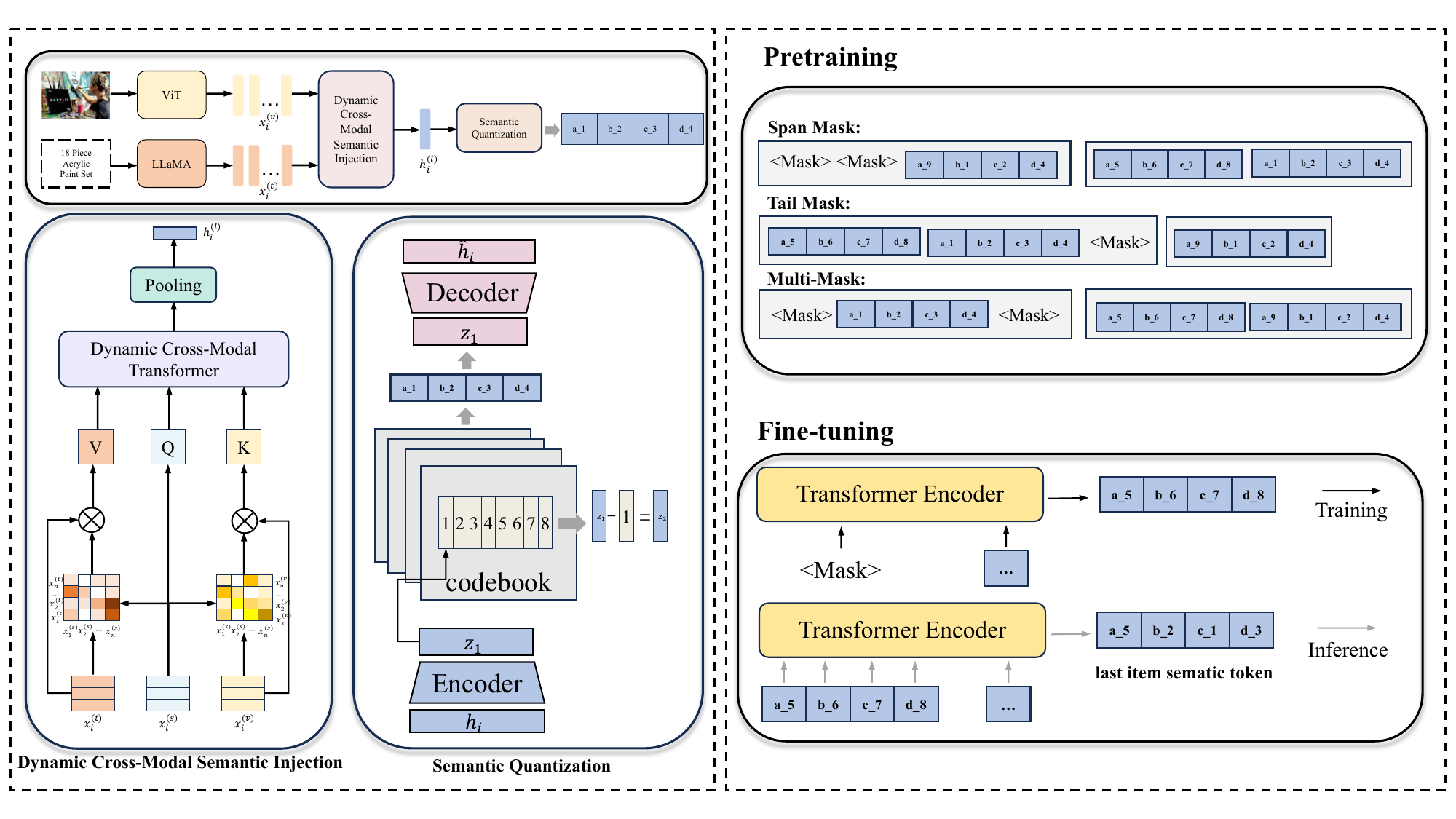}
    \caption{
       An overview of Q-Bert4Rec. Q-Bert4Rec consists of three main stages: Dynamic Cross-Modal Semantic Injection, Semantic Quantization, and Multi-Mask Pretraining and Fine-Tuning.
    }
    \label{fig:layers}
\end{figure*}
\section{RELATED WORK}
\textbf{Sequential recommendation}. Sequential models aims to capture users’ dynamic interaction histories and predict their next potential actions. Early neural architectures, such as GRU4Rec \cite{jannach2017recurrent} and Caser \cite{tang2018personalized}, leveraged recurrent and convolutional structures to capture temporal patterns. Later, the emergence of self-attention mechanisms, particularly SASRec \cite{kang2018self}, greatly improved long-range dependency modeling and parallel training efficiency. Building upon this, BERT4Rec \cite{sun2019bert4rec} introduced a bidirectional Transformer encoder and masked item prediction, setting a strong foundation for subsequent Transformer-based recommenders.Recent works have explored different enhancements on this backbone. For instance, CL4SRec \cite{xie2022contrastive} and DuoRec \cite{qiu2022contrastive} introduced contrastive learning objectives to improve representation robustness, while FDSA \cite{zhang2019feature} incorporated feature-level attention to exploit auxiliary attributes. Despite their success, these approaches still represent items using discrete, non-semantic IDs, which limits their transferability to unseen domains or new items. To address this, a new research trend focuses on \textbf{generative recommendation}, which reformulates recommendation as a language modeling or sequence generation problem. Representative works include P5 \cite{geng2022recommendation}, M6-Rec \cite{cui2022m6}, and TIGER \cite{rajput2023recommender}. These models leverage pre-trained language models (PLMs) and design prompt-based tasks to unify multiple recommendation objectives. P5 and M6-Rec transform recommendation into text generation through task prompts, while P5-ID \cite{hua2023index} explore how the design of item identifiers affects model transferability. ColaRec \cite{wang2024enhanced} constructs generative item IDs by encoding collaborative signals, whereas TIGER\cite{rajput2023recommender} is the first to introduce \textbf{RQ-VAE}\cite{zeghidour2021soundstream} for constructing discrete item IDs via vector quantization. 

\noindent\textbf{Multimodal recommendation}. Multi-modal models exploit side information—such as item text, images, and metadata—to provide richer representations of user–item interactions. 
Early works, such as VBPR \cite{he2016vbpr}, enhanced matrix factorization with visual features extracted from product images. 
Subsequent research introduced graph-based models to capture modality relationships and high-order dependencies. 
For example, MMGCN \cite{wei2019mmgcn} used graph neural networks to propagate multimodal signals; 
LATTICE \cite{zhang2021mining} built modality-specific item-item graphs for structural fusion; 
and DualGNN \cite{wang2021dualgnn} modeled cross-modal attentions between user preferences and item content. 
Similarly, MVGAE \cite{yi2021multi} employed modality-specific variational autoencoders to integrate multimodal embeddings in a unified latent space. With the rapid advancement of large pre-trained models, multimodal recommendation has shifted toward leveraging \textbf{foundation encoders} for representation learning. 
VIP5 \cite{geng2023vip5} extends the P5 framework by incorporating CLIP-based image encoders to align visual and textual modalities within a generative prompting framework. 
MMGRec \cite{liu2024mmgrec} further utilizes a Graph RQ-VAE to construct item IDs from multimodal and collaborative signals, unifying item quantization and relational reasoning. 
IISAN \cite{fu2024iisan} proposes a decoupled PEFT architecture for lightweight multimodal adaptation, demonstrating the scalability of pretrained multimodal encoders. 

\section{METHOD}
\subsection{Problem Formulation}

Let $\mathcal{U}=\{u_1,\dots,u_{N}\}$ be the set of users and $\mathcal{I}=\{i_1,\dots,i_{M}\}$ be the set of items, 
where $N=|\mathcal{U}|$ and $M=|\mathcal{I}|$. 
Each user $u\in\mathcal{U}$ has an interaction sequence 
$\mathcal{S}_u=[\,i_1,i_2,\dots,i_{T_u}\,]$, 
where $i_t\in\mathcal{I}$ is the item interacted at timestamp $t$ and $T_u$ is the sequence length. 
The goal is to predict the next item $\hat{i}_{T_u+1}$ conditioned on the history:
\begin{equation}
\hat{i}_{T_u+1}=\arg\max_{i\in\mathcal{I}} P\!\left(i\,\middle|\,\mathcal{S}_u;\Theta\right),
\end{equation}
where $\Theta$ denotes model parameters.

Unlike traditional ID-based methods, each item $i$ is not represented by an arbitrary integer but by multimodal features 
$\mathbf{x}_i = \{ \mathbf{x}_i^{(t)}, \mathbf{x}_i^{(v)}, \mathbf{x}_i^{(s)} \}$,
where $\mathbf{x}_i^{(t)}$, $\mathbf{x}_i^{(v)}$, and $\mathbf{x}_i^{(s)}$ denote textual, visual, and structural representations, respectively.
Our objective is to learn a quantization mapping 
$f_{\text{quant}} : \mathbf{x}_i \mapsto \mathbf{z}_i$ 
that converts multimodal content into a compact discrete code 
$\mathbf{z}_i = [z_i^1, z_i^2, \dots, z_i^K]$, 
where each $z_i^k \in \{1, 2, \dots, M_c\}$ indexes a semantic token from a codebook $\mathcal{C}$ of size $M_c$. These discrete codes form a \emph{semantic ID} serving as the new modeling unit for sequential recommendation.

Formally, we learn the conditional distribution over the next item given the sequence of semantic IDs:
\begin{equation}
P\!\left(\hat{i}_{T_u+1}\,\middle|\,\mathcal{Z}_u;\Theta\right),\qquad
\mathcal{Z}_u=\big[\,\boldsymbol{z}_{i_1},\boldsymbol{z}_{i_2},\dots,\boldsymbol{z}_{i_{T_u}}\,\big].
\end{equation}
This bridges continuous multimodal semantics and discrete token-level modeling, enabling interpretability.

\subsection{Model Pipeline}
We present the overall architecture of Q-Bert4Rec in Fig. 2.
Q-Bert4Rec is designed as a three-stage framework that unifies multimodal semantic fusion and discrete token modeling for sequential recommendation.
Specifically, it consists of: (1) a dynamic cross-modal semantic injection module that enriches randomly initialized item ID embeddings with textual, visual, and structural semantics through a dynamic transformer; (2) a semantic quantization module that discretizes fused multimodal representations into compact semantic tokens using residual vector quantization (RQ-VAE); and (3) a multi-mask pretraining module that enhances contextual understanding and robustness through diverse masking strategies.

In the first stage, we initialize item IDs as randomly learned embeddings without semantic priors.
To inject multimodal knowledge, each item’s textual, visual, and structural features are projected into a shared embedding space and fused through a dynamic transformer encoder. This module adaptively determines its fusion depth via a learnable gating mechanism, producing semantically enriched ID embeddings that capture cross-modal alignment. In the second stage, the fused embeddings are passed into a residual vector quantization network, which compresses continuous semantic representations into a sequence of discrete tokens. These semantic IDs serve as the new modeling units for downstream recommendation, enabling efficient, interpretable, and transferable item representation learning. In the final stage, we adopt a multi-mask pretraining strategy built upon a BERT-style sequential encoder. Instead of a single random masking scheme, we design three complementary masking objectives—span mask, tail mask, and multi-region mask—to improve the model’s ability to capture both local transition patterns and long-term dependencies.
The pretrained model can then be fine-tuned on domain-specific datasets for next-item prediction or personalized recommendation.

Through these three stages, Q-Bert4Rec effectively bridges the gap between continuous multimodal semantics and discrete item-level reasoning, establishing a unified semantic tokenization pipeline for sequential recommendation.

\subsection{Dynamic Cross-Modal Semantic Injection}

The \textbf{Dynamic Cross-Modal Semantic Injection} module aims to inject multimodal semantics into randomly initialized item IDs, 
enabling the model to align and fuse heterogeneous information from text and images before quantization. 
This module serves as the first stage of Q-Bert4Rec and produces a semantically enriched embedding $\mathbf{h}_i$ for each item.

\noindent\textbf{Input and attention structure.} 
For each item $i$, we maintain a learnable ID vector $\mathbf{q}_i$ as the \emph{query}, 
and extract modality-specific features $\mathbf{x}_i^{(t)}$ and $\mathbf{x}_i^{(v)}$ as the \emph{keys} and \emph{values} 
from pretrained text and image encoders. 
Hence, in our cross-modal attention block, 
$\mathbf{q}_i$ represents the item identity, while $\mathbf{x}_i^{(t)}$ and $\mathbf{x}_i^{(v)}$ 
provide complementary multimodal evidence as key--value pairs.

\noindent\textbf{Similarity-based alignment.}
Before fusion, we perform a lightweight similarity alignment between $\mathbf{q}_i$ and each modality. 
Each modality feature is first normalized and projected into a shared latent space, 
then a cosine similarity is computed to measure its relevance to the ID embedding. 
The similarity scores are converted into soft weights and used to obtain a weighted representation of each modality:
\begin{equation}
\hat{\mathbf{x}}_i^{(m)} = 
\sum_j 
\text{softmax}\!\left(\frac{\mathbf{q}_i^\top \mathbf{x}_{ij}^{(m)}}{\tau}\right) 
\mathbf{x}_{ij}^{(m)}, 
\quad m \in \{t, v\},
\end{equation}
where $\tau$ is a temperature parameter. 
This step allows the model to focus on modality features most semantically relevant to the item ID.

\noindent\textbf{Dynamic cross-modal fusion.}
The aligned textual and visual representations are then fed into a \emph{Dynamic Cross-Modal Transformer} $\mathcal{T}_{dyn}$, 
where $\mathbf{q}_i$ serves as the query and $[\hat{\mathbf{x}}_i^{(t)}; \hat{\mathbf{x}}_i^{(v)}]$ as the key--value inputs. 
Each transformer layer fuses multimodal semantics and updates the query embedding as:
\begin{equation}
\mathbf{h}_i^{(l)} = 
\text{TransformerBlock}(\mathbf{h}_i^{(l-1)}, 
[\hat{\mathbf{x}}_i^{(t)}; \hat{\mathbf{x}}_i^{(v)}]).
\end{equation}
To control the fusion depth adaptively, each layer outputs a \emph{gating vector} 
$\mathbf{g}_i^{(l)} = \sigma(\text{MLP}(\mathbf{h}_i^{(l)}))$, 
where $\sigma(\cdot)$ denotes the sigmoid activation. 
Unlike previous works that use a global scalar or mean value, 
our gating vector is computed through a small MLP and directly determines whether to continue or stop propagation for each item. 
This enables finer-grained control---items with richer semantics are processed through deeper layers, 
while simpler items terminate early.

\noindent\textbf{Training objective.}
To ensure consistent alignment among the multimodal representations, 
we employ a multi-view contrastive loss that jointly optimizes the similarity between the fused representation $\mathbf{h}_i$, 
the modality-aligned features, and the original ID vector. 
The overall loss is defined as:
\begin{equation}
\mathcal{L}_{\text{align}} = 
\mathcal{L}_{it} + \mathcal{L}_{iv} + \mathcal{L}_{tv} ,
\end{equation}
where $\mathcal{L}_{it}$, $\mathcal{L}_{iv}$, and $\mathcal{L}_{tv}$ 
are symmetric InfoNCE losses between the fused, text, and image embeddings, 
The  three terms encourage cross-modal alignment via contrastive learning

The final output $\mathbf{h}_i$ represents the \emph{semantically injected embedding} 
that integrates textual and visual knowledge with the original ID representation. 
It is then used as the input for the following semantic quantization stage.

\subsection{Semantic Quantization}

After obtaining the semantically injected embedding $\mathbf{h}_i$, 
we further convert it into a discrete and compact representation through a \textbf{Residual Vector Quantized Variational Autoencoder (RQ-VAE)}. 
This stage aims to bridge the gap between continuous semantic representations and discrete item identifiers, 
allowing each item to be represented by a set of semantic tokens that preserve multimodal meaning.

\noindent\textbf{Quantization process.} 
Given the semantic embedding $\mathbf{h}_i \in \mathbb{R}^{d}$, 
the encoder of RQ-VAE first projects it into a latent space to produce $\mathbf{z}_i$. 
Then, residual vector quantization discretizes $\mathbf{z}_i$ into $K$ codebooks in a hierarchical manner:
\begin{equation}
\hat{\mathbf{z}}_i = \sum_{k=1}^{K} \mathbf{e}_{q}^{(k)}, 
\quad 
\mathbf{e}_{q}^{(k)} = \arg\min_{\mathbf{e} \in \mathcal{C}^{(k)}} 
\|\mathbf{r}_{i}^{(k-1)} - \mathbf{e}\|_2^2,
\end{equation}
where $\mathcal{C}^{(k)}$ denotes the $k$-th codebook, 
$\mathbf{r}_{i}^{(0)}=\mathbf{z}_i$, and 
$\mathbf{r}_{i}^{(k)}=\mathbf{r}_{i}^{(k-1)}-\mathbf{e}_{q}^{(k)}$ is the residual vector. 
This residual refinement allows RQ-VAE to model fine-grained semantics 
while maintaining compact discrete representations. 
The final quantized code $\hat{\mathbf{z}}_i$ serves as the \emph{semantic ID} for item $i$, 
which replaces the original symbolic identifier.

\noindent\textbf{Reconstruction and loss.} 
The decoder reconstructs the original semantic embedding $\mathbf{h}_i$ from the quantized representation $\hat{\mathbf{z}}_i$. 
The overall training objective consists of two parts: 
a reconstruction loss to preserve semantic information, 
and a quantization loss to stabilize codebook learning.

The reconstruction loss is defined as:
\begin{equation}
\mathcal{L}_{\text{recon}} = 
\|\mathbf{h}_i - \hat{\mathbf{h}}_i\|_2^2,
\end{equation}
which ensures the decoder can faithfully reconstruct the input semantics from the quantized embedding.  

The quantization regularization follows the standard commitment formulation in RQ-VAE:
\begin{equation}
\mathcal{L}_{\text{rq}} = 
\sum_{k=1}^{K}
\Big(
\|\text{sg}[\mathbf{z}_i^{(k)}] - \mathbf{e}_{q}^{(k)}\|_2^2
+ 
\beta \|\mathbf{z}_i^{(k)} - \text{sg}[\mathbf{e}_{q}^{(k)}]\|_2^2
\Big),
\end{equation}
where $\text{sg}[\cdot]$ denotes the stop-gradient operator, 
$\mathbf{e}_{q}^{(k)}$ is the selected codeword in the $k$-th codebook, 
and $\beta$ is the commitment weight controlling codebook stability.  

Finally, the total loss combines the two objectives:
\begin{equation}
\mathcal{L}_{\text{quant}} = 
\mathcal{L}_{\text{recon}} + \mathcal{L}_{\text{rq}},
\end{equation}
This joint optimization ensures both compact codebook utilization and semantic preservation.

\noindent\textbf{Handling collisions.}
In practice, multiple items may be quantized to identical token sequences, 
leading to \emph{code collisions}, where different items have the same token sequences. 
To mitigate this, we adopt the reallocation strategy proposed in MQL4GRec~\cite{zhai2025multimodal}. 
For $N$ colliding items, we first compute the distance tensor 
$\mathbf{D} \in \mathbb{R}^{N \times L \times K}$ 
between residual vectors and codewords in each level:
\begin{equation}
d_i^{(k)} = 
\|\mathbf{r}_i^{(k)} - \mathbf{v}^{(k)}\|_2^2.
\end{equation}
We then sort the distances to obtain the indices $\mathbf{I} = \text{argsort}(\mathbf{D}, \text{axis}=2)$. 
Colliding items are ranked by their minimum distance to the last-level codebook, 
and tokens are reassigned based on the following principles:
(1) Starting from the last codebook, each item receives the nearest available token;  
(2) If collisions persist, reallocation proceeds upward to earlier codebooks in order, 
until all conflicts are resolved.  
This hierarchical reassignment ensures that semantically close items remain nearby 
while preserving code diversity across items.

\noindent\textbf{Output representation.}
Each quantized embedding $\hat{\mathbf{z}}_i$ yields a sequence of discrete indices 
$[z_i^{1}, z_i^{2}, \dots, z_i^{K}]$, 
each corresponding to one codeword in the learned codebooks. 
To form a unified vocabulary for downstream recommendation, 
we serialize these indices into symbolic tokens by prefixing a lowercase identifier for each codebook, 
e.g.,\texttt{<a\_2><b\_3><c\_1><d\_6>}, where \texttt{a–d} denote codebook identifiers and the subscripts represent token indices. This compact and interpretable tokenization enables our model to replace arbitrary item IDs with meaningful \emph{semantic IDs}, which serve as the fundamental modeling units for subsequent sequential recommendation and pretraining stages.

\subsection{Multi-Mask Pretrain}

After obtaining the quantized semantic IDs, we perform pretraining on user interaction sequences to enhance the model’s ability to understand multimodal semantics and temporal dependencies.  
Different from conventional BERT4Rec-style objectives that rely on a single random masking policy, we introduce a \textbf{Multi-Mask Pretraining} strategy, which leverages three complementary masking schemes—span, tail, and multi-region masking—to capture both short-term transitions and long-range contextual relations.

\noindent\textbf{Masking strategies.}
To enrich the pretraining signal and improve robustness, 
we design three masking schemes that are jointly sampled during training: (1)\textbf{Span masking.} A consecutive segment of tokens is masked, encouraging the model to model local coherence and short-term item transitions within user sessions.(2)\textbf{Tail masking.} The last few tokens in the sequence are masked to simulate next-item prediction, enabling the model to anticipate future interactions based on historical behavior.(3)\textbf{Multi-region masking.} Several non-contiguous regions are randomly masked, requiring the model to infer missing tokens from both nearby and distant contexts, thereby enhancing its long-range reasoning capacity.

\subsection{Training and Fine-tuning}

\textbf{Training.}
Our model follows a two-stage paradigm of \emph{pretraining} and \emph{fine-tuning}.  
During \textbf{pretraining}, 
we utilize large-scale multi-domain datasets to learn general recommendation priors.  
This stage enables the model to capture cross-domain sequential patterns and multimodal semantics, 
serving as a foundation for downstream personalization.
During \textbf{fine-tuning}, 
we adapt the pretrained model to the target-domain dataset using the standard masked prediction objective.  
Object of two stage is the same, given a user’s semantic interaction sequence 
$\mathcal{Z}_u = [\mathbf{z}_{i_1}, \mathbf{z}_{i_2}, \dots, \mathbf{z}_{i_{T_u}}]$, 
a subset of tokens is randomly masked, and the model is trained to recover the masked tokens based on their surrounding context:
\begin{equation}
\mathcal{L}_{\text{mask}} 
= - \sum_{t \in \mathcal{M}} 
\log P_{\theta}(\mathbf{z}_{i_t} \mid \mathcal{Z}_u^{\backslash \mathcal{M}}),
\end{equation}
where $\mathcal{M}$ denotes the set of masked positions and 
$\mathcal{Z}_u^{\backslash \mathcal{M}}$ represents the unmasked sequence.  
This objective encourages the model to understand temporal dependencies and user preference evolution within a sequence.

\noindent\textbf{Inference.}
At inference time, 
we mask the last interaction in a user’s sequence and predict the most probable semantic ID as the next item:
$\hat{i}_{T_u+1} = 
\arg\max_{i \in \mathcal{I}} 
P_{\theta}(\mathbf{z}_{i} \mid 
\mathbf{z}_{i_1}, \dots, \mathbf{z}_{i_{T_u-1}})$.
This simple yet effective procedure allows the model to leverage the pretrained multimodal priors and domain-specific sequential knowledge 
for accurate next-item recommendation.
\begin{table}[t]
\centering
\caption{Statistics of the preprocessed datasets. 
}
\label{tab:dataset_stat}
\resizebox{\linewidth}{!}{
\begin{tabular}{lccccc}
\toprule
\textbf{Datasets} & \textbf{\#Users} & \textbf{\#Items} & \textbf{\#Interactions} & \textbf{Sparsity} & \textbf{Avg.\ len} \\
\midrule
Pet          & 183,697 & 31,986 & 1,571,284 & 99.97\% & 8.55 \\
Cell         & 123,885 & 38,298 &   873,966 & 99.98\% & 7.05 \\
Automotive   & 105,490 & 39,537 &   845,454 & 99.98\% & 8.01 \\
Tools        & 144,326 & 41,482 & 1,153,959 & 99.98\% & 8.00 \\
Toys         & 135,748 & 47,520 & 1,158,602 & 99.98\% & 8.53 \\
Sports       & 191,920 & 56,395 & 1,504,646 & 99.99\% & 7.84 \\
\midrule
Instruments  & 17,112  &  6,250 &   136,226 & 99.87\% & 7.96 \\
Arts         & 22,171  &  9,416 &   174,079 & 99.92\% & 7.85 \\
Games        & 42,259  & 13,839 &   373,514 & 99.94\% & 8.84 \\
\bottomrule
\end{tabular}
}
\end{table}

\begin{table*}[t]
\centering
\caption{Performance comparison on three Amazon sub-datasets with grouped column headers.
Best results are in \textbf{bold}; second best are \underline{underlined}.
Improv. denotes the relative improvement of \textbf{Ours} over best model.}
\setlength{\tabcolsep}{3.8pt}
\resizebox{\linewidth}{!}{
\begin{tabular}{c l c c c c c c c c c c c c c}
\toprule
\textbf{Dataset} & \textbf{Metric}
& \multicolumn{1}{c}{\textit{Traditional}}
& \multicolumn{4}{c}{\textit{Transformer-based}}
& \multicolumn{2}{c}{\textit{VQ/Multimodal}}
& \multicolumn{2}{c}{\textit{Prompt-LLM}}
& \multicolumn{2}{c}{\textit{Semantic-ID / Gen.}}
& \multicolumn{1}{c}{}
& \multicolumn{1}{c}{}
\\
\cmidrule(lr){3-3}\cmidrule(lr){4-7}\cmidrule(lr){8-9}\cmidrule(lr){10-11}\cmidrule(lr){12-13}
 &  &
 GRU4Rec &
 BERT4Rec & SASRec & FDSA & S$^3$-Rec &
 VQ-Rec & MISSRec &
 P5\text{-}CID & VIP5 &
 TIGER & MQL4GRec &
 \textbf{Ours} & \textbf{Improv.}
\\
\midrule
\multirow{5}{*}{Instrument} & HR@1    & 0.0566 & 0.0450 & 0.0318 & 0.0530 & 0.0339 & 0.0502 & 0.0723 & 0.0512 & 0.0737 & 0.0754 & \underline{0.0833} & \textbf{0.0835} & +0.24\% \\
 & HR@5    & 0.0975 & 0.0856 & 0.0946 & 0.0987 & 0.0937 & 0.1062 & 0.1089 & 0.0839 & 0.0892 & 0.1007 & \underline{0.1115} & \textbf{0.1153} & +3.41\% \\
 & HR@10   & 0.1207 & 0.1081 & 0.1233 & 0.1249 & 0.1123 & 0.1357 & 0.1361 & 0.1119 & 0.1071 & 0.1221 & \underline{0.1375} & \textbf{0.1418} & +3.13\% \\
 & NDCG@5  & 0.0783 & 0.0667 & 0.0654 & 0.0775 & 0.0693 & 0.0796 & 0.0797 & 0.0678 & 0.0815 & 0.0882 & \underline{0.0977} & \textbf{0.0995} & +1.84\% \\
 & NDCG@10 & 0.0857 & 0.0739 & 0.0746 & 0.0859 & 0.0743 & 0.0891 & 0.0880 & 0.0704 & 0.0872 & 0.0950 & \underline{0.1060} & \textbf{0.1077} & +1.60\% \\
\midrule
\multirow{5}{*}{Arts} & HR@1    & 0.0365 & 0.0289 & 0.0212 & 0.0380 & 0.0172 & 0.0408 & 0.0479 & 0.0421 & 0.0474 & 0.0532 & \underline{0.0672} & \textbf{0.0756} & +12.50\% \\
 & HR@5    & 0.0817 & 0.0697 & 0.0951 & 0.0832 & 0.0739 & \underline{0.1038} & 0.1021 & 0.0713 & 0.0704 & 0.0894 & 0.1037 & \textbf{0.1136} & +9.44\% \\
 & HR@10   & 0.1088 & 0.0922 & 0.1250 & 0.1190 & 0.1030 & \underline{0.1386} & 0.1321 & 0.0994 & 0.0859 & 0.1167 & 0.1327 & \textbf{0.1419} & +2.38\% \\
 & NDCG@5  & 0.0602 & 0.0502 & 0.0610 & 0.0583 & 0.0511 & 0.0732 & 0.0699 & 0.0607 & 0.0586 & 0.0718 & \underline{0.0857} & \textbf{0.0947} & +10.50\% \\
 & NDCG@10 & 0.0690 & 0.0575 & 0.0706 & 0.0695 & 0.0630 & 0.0844 & 0.0815 & 0.0662 & 0.0635 & 0.0806 & \underline{0.0950} & \textbf{0.1039} & +9.36\% \\
\midrule
\multirow{5}{*}{Games} & HR@1    & 0.0140 & 0.0115 & 0.0069 & 0.0163 & 0.0136 & 0.0075 & 0.0201 & 0.0169 & 0.0173 & 0.0166 & \underline{0.0203} & \textbf{0.0233} & +14.77\% \\
 & HR@5    & 0.0544 & 0.0426 & 0.0587 & 0.0614 & 0.0527 & 0.0408 & \textbf{0.0674} & 0.0532 & 0.0480 & 0.0523 & 0.0637 & \underline{0.0673} & - \\
 & HR@10   & 0.0895 & 0.0725 & 0.0985 & 0.0998 & 0.0903 & 0.0679 & \underline{0.1048} & 0.0930 & 0.0758 & 0.0857 & 0.1033 & \textbf{0.1061} & +1.24\% \\
 & NDCG@5  & 0.0341 & 0.0270 & 0.0333 & 0.0389 & 0.0351 & \underline{0.0422} & 0.0385 & 0.0334 & 0.0328 & 0.0345 & 0.0421 & \textbf{0.0451} & +6.87\% \\
 & NDCG@10 & 0.0453 & 0.0366 & 0.0461 & 0.0509 & 0.0468 & 0.0329 & 0.0499 & 0.0454 & 0.0418 & 0.0453 & \underline{0.0548} & \textbf{0.0575} & +4.93\% \\
\bottomrule
\end{tabular}
}
\label{tab:main_results}
\end{table*}


\begin{table*}[t]
\centering
\caption{Ablation study on modality contribution. \cmark\ indicates the enabled modality.}
\resizebox{\textwidth}{!}{
\begin{tabular}{ccccccccccccccc}
\toprule
\multicolumn{3}{c}{\textbf{Modality}} &
\multicolumn{4}{c}{\textbf{Instruments}} &
\multicolumn{4}{c}{\textbf{Arts}} &
\multicolumn{4}{c}{\textbf{Games}} \\ 
\cmidrule(lr){1-3} \cmidrule(lr){4-7} \cmidrule(lr){8-11} \cmidrule(lr){12-15}
\textbf{Image} & \textbf{Text} & \textbf{ID-Only} &
HR@5 & HR@10 & NDCG@5 & NDCG@10 &
HR@5 & HR@10 & NDCG@5 & NDCG@10 &
HR@5 & HR@10 & NDCG@5 & NDCG@10 \\
\midrule
\cmark &  &  & 0.1126 & 0.1367 & 0.0977 & 0.1055 & 0.1112 & 0.1399 & 0.0928 & 0.1020 & 0.0660 & 0.1048 & 0.0441 & 0.0566 \\
 & \cmark &  & 0.1139 & 0.1398 & 0.0979 & 0.1063 & 0.1111 & 0.1378 & 0.0922 & 0.1007 & 0.0646 & 0.1040 & 0.0437 & 0.0563 \\
 &  & \cmark & 0.1117 & 0.1363 & 0.0955 & 0.1034 & 0.1079 & 0.1367 & 0.0890 & 0.0982 & 0.0625 & 0.1011 & 0.0415 & 0.0539 \\
\cmark & \cmark & \cmark & \textbf{0.1153} & \textbf{0.1418} & \textbf{0.0995} & \textbf{0.1077} & \textbf{0.1136} & \textbf{0.1419} & \textbf{0.0947} & \textbf{0.1039} & \textbf{0.0673} & \textbf{0.1061} & \textbf{0.0451} & \textbf{0.0575} \\
\bottomrule
\end{tabular}
}
\label{tab:ablation_modality_full}
\end{table*}

\section{Experiments}
To validate the effectiveness of our proposed Q-BERT4Rec method, we conduct a lot of experiments and answer the following research questions:

\noindent\textbf{RQ1:} How does Q-BERT4Rec perform compared with existing best-performing recommendation models among different datasets?\\
\textbf{RQ2:} How do the three stages of Q-BERT4Rec, namely dynamic cross-modal semantic injection, semantic quantization, and multi-mask pretraining, affect the performance of our model?\\
\textbf{RQ3:} How do different hyper-parameter settings affect the performance of Q-BERT4Rec?

\noindent\textbf{RQ4:} Does dynamic cross-modal semantic injection better than tranditional multi-modal fusion?
\subsection{Experimental Setting}
\textbf{Dataset}. We conduct experiments on the widely used \textbf{Amazon Product Reviews} dataset\cite{ni2019justifying}, 
which contains user–item interaction histories of different domains collected from Amazon over several years.  
Each domain represents a distinct product category and includes user reviews, ratings, and timestamps, 
making it suitable for evaluating cross-domain sequential recommendation.Statistics of these datasets are shown in Table ~\ref{tab:dataset_stat}.

To verify the transferability and generalization ability of our proposed Q-BERT4Rec framework, 
we divide the dataset into two parts for \textbf{pretraining} and \textbf{fine-tuning}.  
Specifically, we use six source domains for pretraining:
\textit{Pet Supplies}, 
\textit{Cell Phones and Accessories}, 
\textit{Automotive}, 
\textit{Tools and Home Improvement}, 
\textit{Toys and Games}, and 
\textit{Sports and Outdoors}.  
These domains provide rich and diverse behavioral sequences that help the model learn general recommendation priors.  

For \textbf{fine-tuning}, 
we adopt three target domains with distinct user behaviors and item distributions:  
\textit{Musical Instruments}, 
\textit{Arts, Crafts and Sewing}, and 
\textit{Video Games}.  
This setting evaluates the model’s ability to transfer multimodal and sequential knowledge 
from heterogeneous source domains to new target domains.

\noindent\textbf{Evaluation Metrics}. We evaluate model performance using two widely adopted ranking metrics:  
Hit Ratio (HR@K) and Normalized Discounted Cumulative Gain (NDCG@K) with K = 1, 5, 10. Following previous work\cite{kang2018self}, We use the \textit{leave-one-out} for evaluation, where the most recent interaction is held out for testing and the second most recent for validation.

\noindent\textbf{Implementation Details.} 
We implement our model using \texttt{PyTorch}. 
For the dynamic cross-modal semantic injection module, 
we employ the pretrained \textbf{LLaMA} model to encode textual representations, 
and adopt the image encoder branch from \textbf{CLIP}~\cite{radford2021learning}, 
which uses \textbf{ViT}~\cite{dosovitskiy2020image} as its backbone, 
to extract visual features of items. 
The Transformer within this module is configured with a maximum depth of 5 layers. 

Both the encoder and decoder in the \textbf{RQ-VAE} module are implemented as multilayer perceptrons (MLPs) with \texttt{ReLU} activation functions. 
We set the number of codebook levels to 4, 
each containing 256 vectors with a dimensionality of 32. 
The learning rate and batch size are set to 0.001 and 1024, respectively. 

Following prior work~\cite{sun2019bert4rec}, 
during pretraining, we apply three masking strategies with probabilities of 0.3 (span mask), 0.15 (multi-position mask), and 0.1 (tail mask). 
The model is trained with a batch size of 512 and a learning rate of 0.0005. 
The Transformer encoder consists of 4 layers with 4 attention heads. 
For fine-tuning, we use only the traditional mask strategy with a ratio of 0.3, 
a batch size of 256, and a learning rate of 0.0001. 
To prevent overfitting, we apply dropout with a rate of 0.2 to all Transformer layers. 

All experiments are conducted on NVIDIA RTX 4090 GPUs. 
We use four GPUs during pretraining and a single GPU for fine-tuning.
\begin{table}[t]
\centering
\caption{Ablation study on pretraining strategies. We compare without pretraining, traditional MLM pretraining, and our multi-mask pretraining.}
\resizebox{\linewidth}{!}{
\begin{tabular}{llcccccc}
\toprule
\textbf{Modal} & \textbf{Tasks} &
\multicolumn{2}{c}{\textbf{Instruments}} &
\multicolumn{2}{c}{\textbf{Arts}} &
\multicolumn{2}{c}{\textbf{Games}} \\ 
\cmidrule(lr){3-4} \cmidrule(lr){5-6} \cmidrule(lr){7-8}
 &  & HR@10 & NDCG@10 & HR@10 & NDCG@10 & HR@10 & NDCG@10 \\
\midrule
\multirow{3}{*}{All} 
& w/o Pretrain & 0.1360 & 0.1035 & 0.1361 & 0.0988 & 0.1037 & 0.0543 \\
& + MLM        & 0.1384 & 0.1055 & 0.1405 & 0.1019 & 0.1041 & 0.0566 \\
& + MultiMask  & \textbf{0.1418} & \textbf{0.1077} & \textbf{0.1419} & \textbf{0.1039} & \textbf{0.1061} & \textbf{0.0575} \\
\bottomrule
\end{tabular}
}
\label{tab:ablation_pretrain}
\end{table}

\subsection{Performance Comparison (RQ1)}
\textbf{Baselines}. In this section, we compare our proposed approach with a wide range of sequential recommendation baselines, which are briefly described in Appendix~\ref{appendix:baseline}.
The compared methods include GRU4Rec\cite{jannach2017recurrent}, BERT4Rec \cite{sun2019bert4rec}, SASRec\cite{kang2018self}, FDSA\cite{zhang2019feature}, S³-Rec\cite{zhou2020s3}, VQ-Rec\cite{hou2023learning}, MISSRec\cite{wang2023missrec}, P5-CID\cite{hua2023index}, VIP5\cite{geng2023vip5}, TIGER\cite{rajput2023recommender}, and MQL4Grec\cite{zhai2025multimodal}. The experimental results are reported in Table~\ref{tab:main_results}.

\noindent\textbf{Results and Analysis.}  
As shown in Table~\ref{tab:main_results}, our proposed Q-BERT4Rec consistently outperforms all baselines across three downstream datasets, including \textit{Musical Instruments}, \textit{Arts, Crafts \& Sewing}, and \textit{Video Games}.  
The improvement is evident across both ranking metrics (HR@$K$ and NDCG@$K$), demonstrating the effectiveness of integrating semantic tokenization with multimodal fusion.

On the \textit{Instrument} dataset, Q-BERT4Rec achieves an HR@5 of 0.1153 and NDCG@10 of 0.1077, outperforming the strongest baseline MQL4GRec by +3.41\% and +1.60\%, respectively.  
For the \textit{Arts} dataset, our model achieves 0.1419 in HR@10 and 0.1039 in NDCG@10, yielding relative gains of +2.38\% and +9.36\% compared with the previous best model.  
The largest improvement is observed on the \textit{Games} dataset, where Q-BERT4Rec surpasses MQL4GRec by +14.77\% in HR@1 and +6.87\% in NDCG@5, highlighting its robustness in sparse and noisy interaction scenarios.

These consistent improvements validate that dynamic cross-modal semantic injection effectively captures richer multimodal dependencies, while the residual quantization module enhances semantic compactness and generalization across domains.  
Furthermore, the multi-mask pretraining strategy strengthens the model’s ability to understand diverse user interaction patterns, leading to better next-item prediction accuracy.  
Overall,in answer to RQ1, Q-BERT4Rec achieves the best performance in 13 out of 15 evaluation metrics, demonstrating strong generalization, and robustness across heterogeneous recommendation domains.
\vspace{-0.5em}
\subsection{Ablation Study (RQ2)}
To evaluate the effectiveness of each component in Q-BERT4Rec, 
we conduct ablation experiments from two aspects: 
(\textit{i}) the contribution of different modalities in the dynamic cross-modal semantic injection module, 
and (\textit{ii}) the influence of various pretraining strategies. 
The detailed results are reported in Table~\ref{tab:ablation_modality_full} and Table~\ref{tab:ablation_pretrain}.

\noindent\textbf{Impact of modality.}
As shown in Table~\ref{tab:ablation_modality_full}, 
removing any modality leads to a consistent degradation in performance across all datasets. 
For instance, on the \textit{Instruments} dataset, 
the model using only image features achieves HR@10 of 0.1367 and NDCG@10 of 0.1055, 
while the text-only variant reaches HR@10 of 0.1398 and NDCG@10 of 0.1063. 
However, when all three modalities (text, image, and ID) are jointly utilized, 
the performance improves to HR@10 of \textbf{0.1419} and NDCG@10 of \textbf{0.1077}, 
demonstrating a clear synergistic effect among modalities. 
A similar trend is also observed on the \textit{Arts} and \textit{Games} datasets.
These results confirm that the integration of visual, textual, and structural cues 
provides complementary information that enables richer semantic representations of user-item interactions.
\begin{figure}[t]
    \centering
    \includegraphics[width=0.9\linewidth]{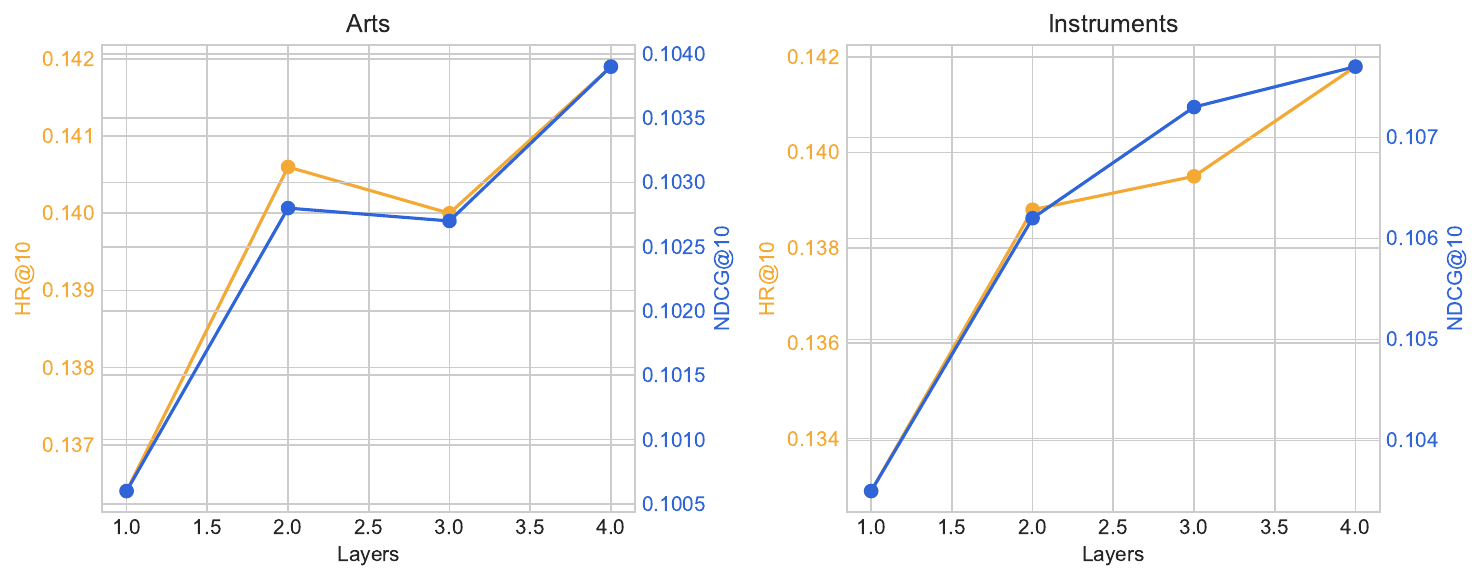}
    \caption{
        Analysis of the number of transformer layers
    }
    \label{fig:layers}
\end{figure}

\noindent\textbf{Impact of pretraining strategies.}
Table~\ref{tab:ablation_pretrain} presents the effect of different pretraining schemes. 
Without pretraining, the model performs relatively poorly 
(e.g., HR@10 = 0.1360, NDCG@10 = 0.1035 on \textit{Instruments}), 
showing that learning from scratch limits the model’s ability to capture sequential dependencies. 
Introducing traditional MLM pretraining improves performance to HR@10 = 0.1384 and NDCG@10 = 0.1055, 
highlighting the benefit of contextual reconstruction. 
Furthermore, our proposed \textit{multi-mask pretraining} achieves the best performance across all datasets, 
reaching HR@10 = \textbf{0.1419} and NDCG@10 = \textbf{0.1077} on \textit{Instruments}, 
HR@10 = \textbf{0.1419} and NDCG@10 = \textbf{0.1039} on \textit{Arts}, 
and HR@10 = \textbf{0.1061} and NDCG@10 = \textbf{0.0575} on \textit{Games}. 
Compared with the MLM-only variant, multi-mask pretraining brings an average gain of 
\textbf{+2.27\% in HR@10} and \textbf{+1.70\% in NDCG@10}, 
indicating that combining span, tail, and multi-region masking helps the model 
learn more robust sequence representations and adapt better to different user behaviors.

\noindent\textbf{Summary.}
Overall, to answer the questions of RQ2, these ablation studies validate that both multimodal alignment and 
multi-mask pretraining are indispensable to Q-BERT4Rec’s success. 
The former injects rich semantic knowledge into item representations, 
while the latter enhances the model’s sequential understanding, 
together enabling stronger generalization and recommendation accuracy.
\subsection{Hyper-Parameter Analysis (RQ3)}
We further investigate the sensitivity of Q-BERT4Rec to key hyperparameters that influence model performance and generalization. 
Specifically, we analyze the effects of three important parameters: 
(\textit{i}) the number of Transformer layers, 
(\textit{ii}) the masking probability used during finetuning, 
and (\textit{iii}) the dropout rate applied to Transformer layers. 
We report the recommendation performance in terms of HR@10, 
as we observe similar trends for NDCG@10 across all datasets. 
For clarity, Figure~\ref{fig:layers}--\ref{fig:mask} present the results on the \textit{Arts} and \textit{Instruments} datasets, 
as other domains show consistent patterns. 
By varying these hyperparameters, we aim to understand their impact on both model expressiveness and generalization ability.

\noindent\textbf{Impact of Layer Number}. To analyze the effect of model depth, we vary the number of Transformer layers from 1 to 4 and report the performance on the Arts and Instruments datasets, as illustrated in Figure~\ref{fig:layers} . The results show that increasing the number of layers consistently improves both HR@10 and NDCG@10 in the early stages, indicating that deeper architectures enhance the model’s ability to capture sequential dependencies and cross-modal interactions.
For instance, in Instruments, HR@10 rises from 0.1329 (1 layer) to 0.1418 (4 layers), while NDCG@10 improves from 0.1035 to 0.1077. A similar trend can be observed in Arts, where HR@10 increases from 0.1364 to 0.1419 and NDCG@10 improves from 0.1006 to 0.1039. However, the deeper the layers, the slower the training and inference speed.

\begin{figure}[t]
    \centering
    \includegraphics[width=0.9\linewidth]{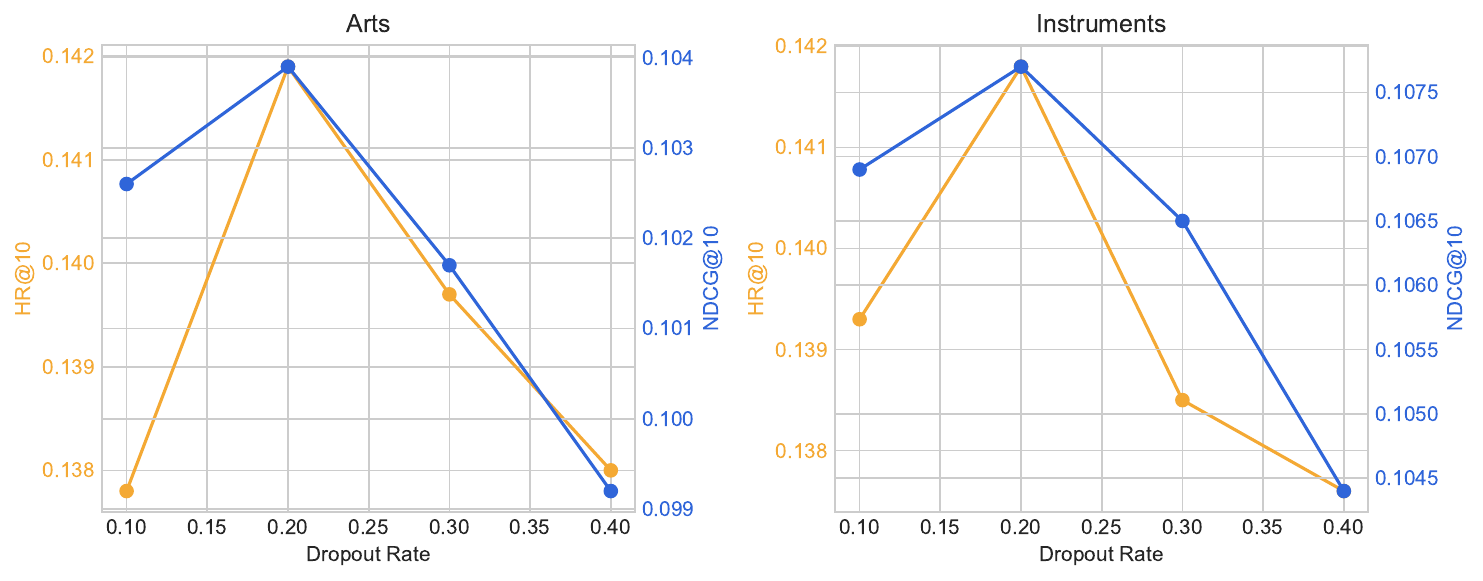}
    \caption{
        Analysis of the impact of different dropout rates
    }
    \label{fig:dropout}
\end{figure}

\begin{figure}[t]
    \centering
    \includegraphics[width=0.9\linewidth]{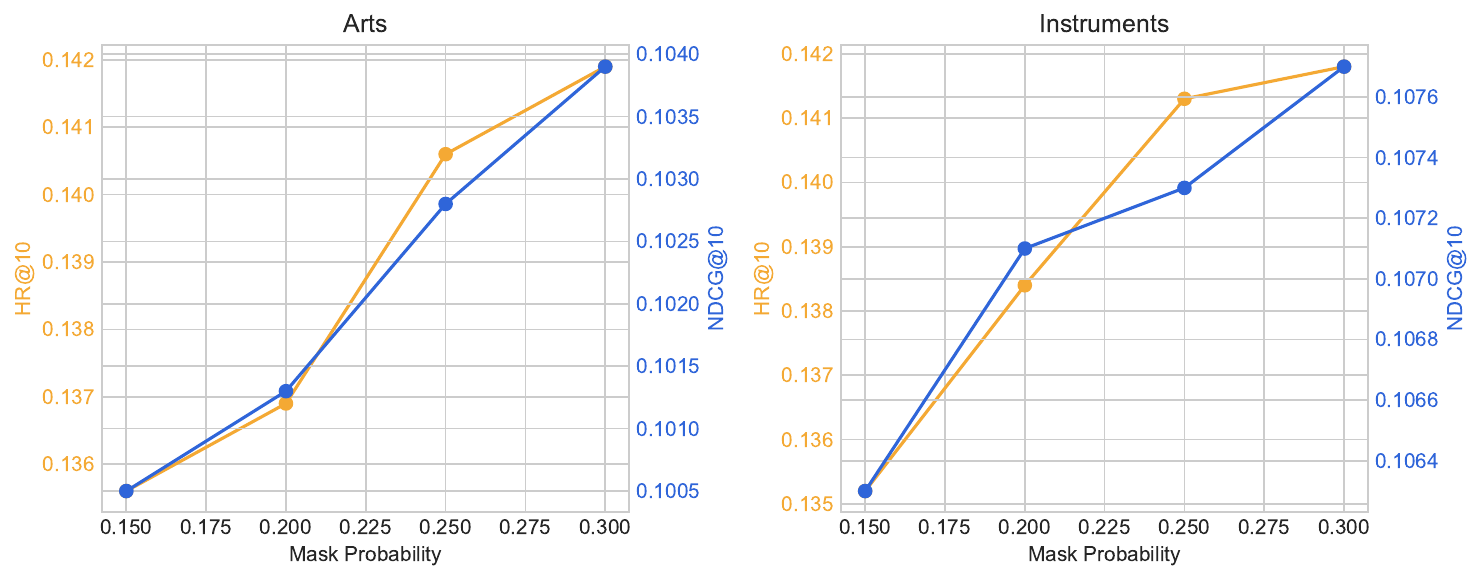}
    \caption{
        Analysis of the impact of different mask probabilities
    }
    \label{fig:mask}
\end{figure}

\noindent\textbf{Impact of Dropout Rate}. To further investigate the generalization ability of our model, we also analyze the impact of different dropout rates on performance, as illustrated in Figure~\ref{fig:dropout}.
We vary the dropout rate from 0.1 to 0.4 and observe its effect on HR@10 and NDCG@10 across the Arts and Instruments datasets. As shown in the figure, performance first increases and then gradually declines as the dropout rate grows. A moderate dropout rate (around 0.2) achieves the best results, improving HR@10 to 0.1419 and 0.1418 on Arts and Instruments, respectively. This demonstrates that a small amount of regularization effectively prevents overfitting and stabilizes training, while excessive dropout (e.g., 0.4) leads to insufficient feature utilization and degraded recommendation accuracy.
Therefore, we adopt 0.2 as the optimal dropout setting in all subsequent experiments to balance regularization and representation learning.

\noindent\textbf{Impact of Mask Probability}. We further investigate how the masking probability affects the model’s pretraining performance, as illustrated in Figure~\ref{fig:mask}. We vary the mask probability from 0.15 to 0.30 on the Arts and Instruments datasets and report the corresponding HR@10 and NDCG@10 values.
As shown in the figure, both metrics exhibit a clear upward trend as the mask ratio increases, indicating that a higher masking probability encourages the model to capture more contextual dependencies within user behavior sequences. However, the improvement plateaus when the probability exceeds 0.3, suggesting that excessive masking may remove too much information, leading to suboptimal reconstruction and weaker sequential understanding. The best performance is achieved when the mask probability is set to 0.3, where HR@10 reaches 0.1419 on Arts and 0.1418 on Instruments. Therefore, we adopt 0.3 as the default masking ratio in all subsequent experiments to balance information preservation and representation learning.

\subsection{Modality Fusion of Q-Bert4Rec (RQ4)}
\begin{figure}[t]
    \centering
    \includegraphics[width=0.9\linewidth, trim=0 100 0 100,clip]{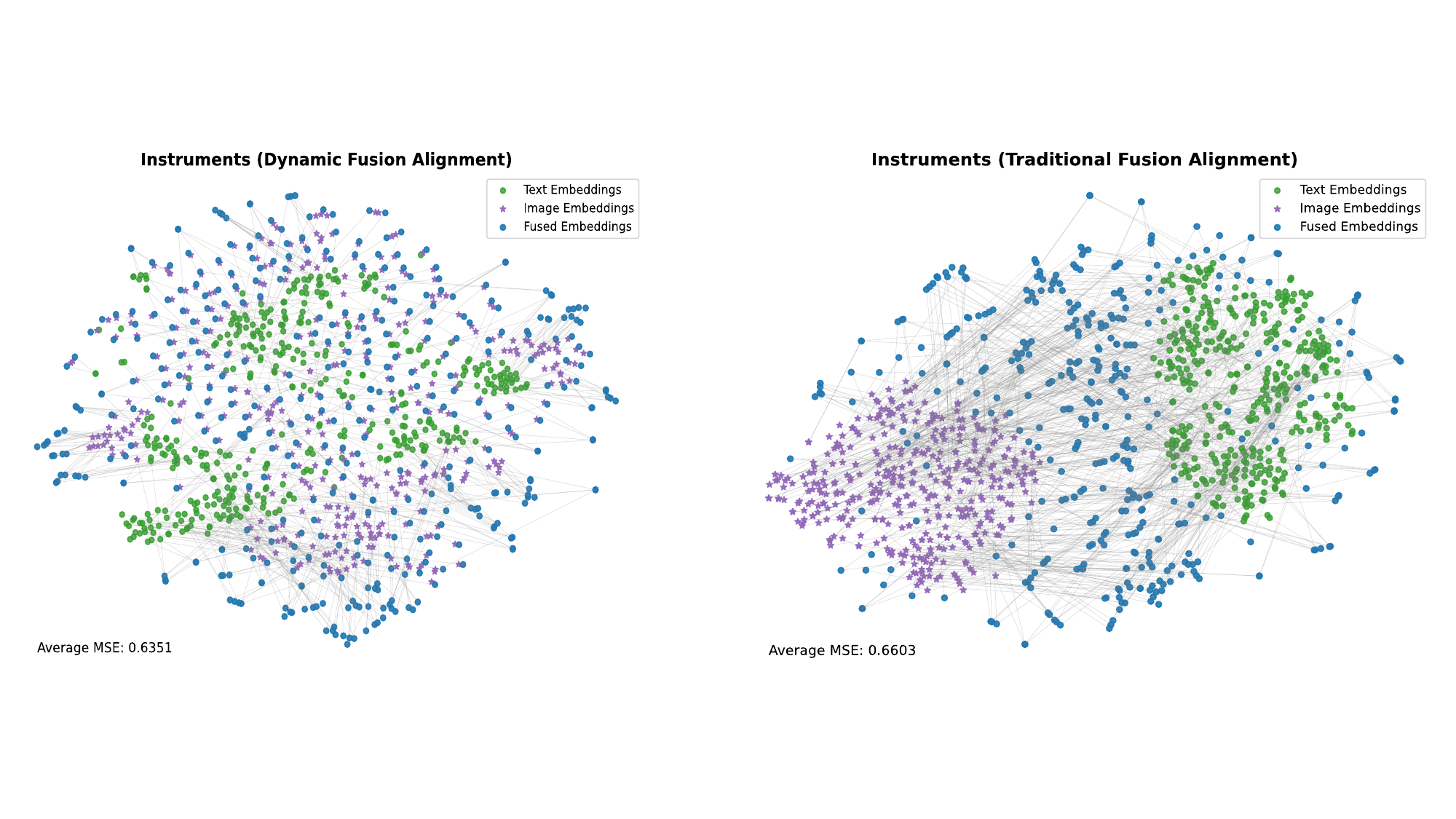}
    \caption{
        Dynamic Modality Fusion compare with traditional method.
    }
    \label{fig:fusion}
\end{figure}

\begin{figure}[t]
    \centering
    \includegraphics[width=0.45\textwidth, trim=200 130 200 110,clip]{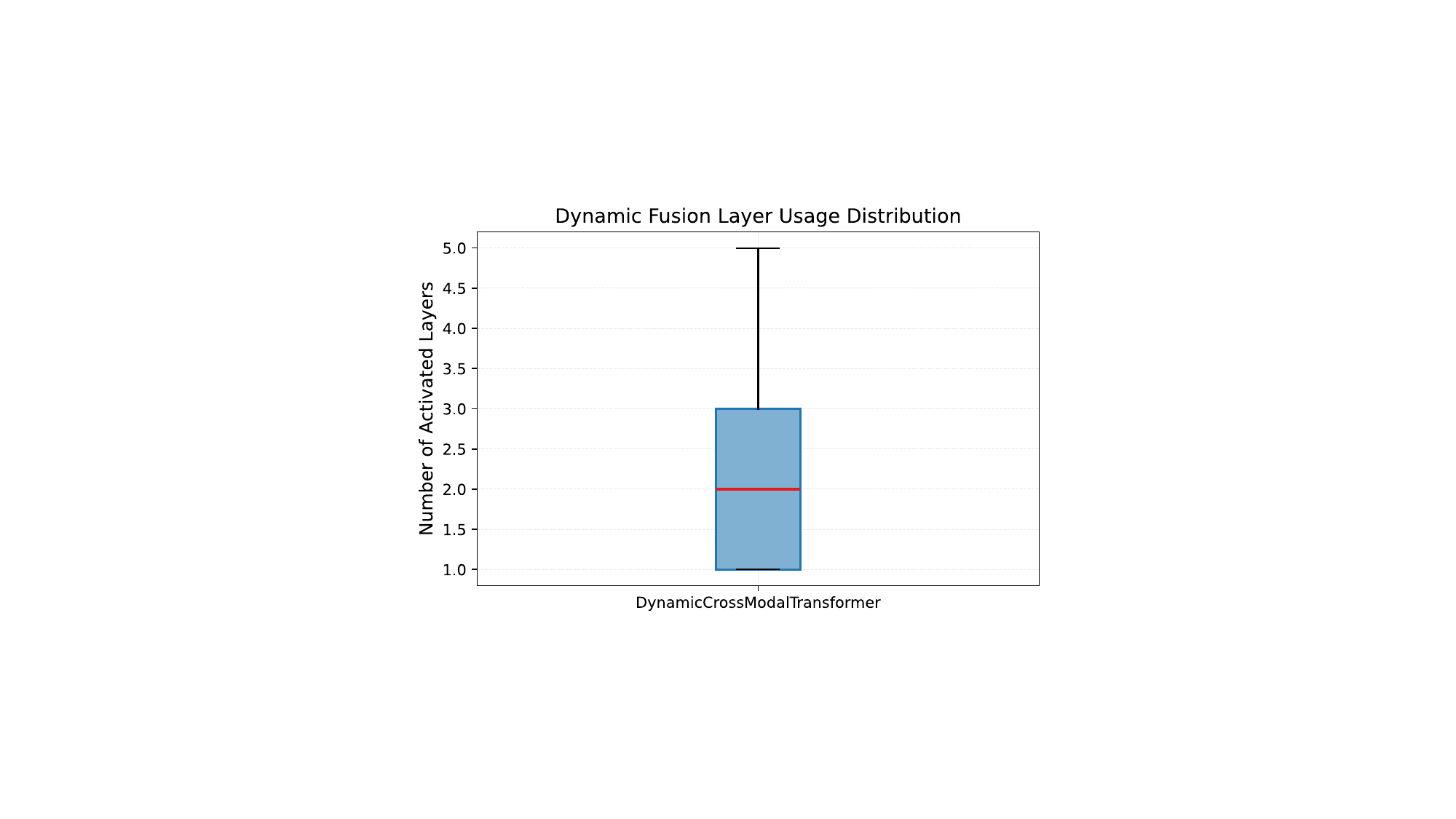}
    \caption{
        Distribution of activated layers in the Dynamic Cross-Modal Transformer.
    }
    \label{fig:box}
\end{figure}

To better understand how our Dynamic Fusion Transformer aligns multi-modal representations,
we visualize the learned textual, visual, and fused embeddings on the Instruments dataset.
Each item is represented by three points — textual (green), visual (purple), and fused (blue) —
and connected by gray lines that indicate their pairwise relations in the semantic space.
As shown in Figure~\ref{fig:fusion} (left), the proposed Dynamic Fusion Alignment yields much tighter and more coherent clusters,
where the fused embeddings consistently lie between textual and visual modalities, forming semantically meaningful triplets.
In contrast, Traditional Fusion which is fixed the layer number (Figure~\ref{fig:fusion} , right) exhibits clearly separated modality-specific regions,
indicating weak cross-modal consistency and the absence of adaptive semantic bridging.
The lower average MSE (0.6351 vs. 0.6603) demonstrates that dynamic fusion effectively narrows the modality gap while maintaining complementary information from each modality. Furthermore, Figure~\ref{fig:box} shows the distribution of activated layers in the Dynamic Cross-Modal Transformer. Most samples use between one and three fusion layers, with a median depth of two. This observation suggests that the model adaptively adjusts its fusion depth according to modality complexity — simpler samples terminate early, while complex ones traverse deeper layers.

\section{Conclusion and Future Work}

In this work, we introduced a semantic-ID based sequential recommendation framework that unifies dynamic cross-modal fusion with residual quantization. The proposed dynamic fusion module adaptively selects the depth of cross-modal interaction, enabling the model to effectively integrate heterogeneous textual and visual semantics. Building upon these fused representations, our RQ-VAE module transforms item features into compact and transferable semantic IDs, which significantly enhance sequence modeling and cross-domain generalization. Extensive experiments across three real-world domains demonstrate consistent performance gains over strong multimodal and generative baselines. In future work, we plan to expand semantic-ID modeling to larger-scale catalogs and more modalities, and explore its potential in generative recommendation and unified multimodal pretraining.

\bibliographystyle{ACM-Reference-Format}
\bibliography{ref}
\clearpage


\appendix
\section*{Appendix}
\section{Baselines}
\label{appendix:baseline}
\noindent\textbf{(1) Traditional Models.}  
\begin{itemize}[leftmargin=*]
  \item \textbf{GRU4Rec}~\cite{jannach2017recurrent} employs gated recurrent units to model sequential dependencies in user behaviors.  
  Although effective for short sequences, it struggles to capture long-range dependencies and contextual semantics.
\end{itemize}

\noindent\textbf{(2) Transformer-based Models.}  
\begin{itemize}[leftmargin=*]
  \item \textbf{BERT4Rec}~\cite{sun2019bert4rec} applies bidirectional Transformers with the Masked Language Modeling (MLM) objective for next-item prediction.  
  However, it models discrete item IDs without semantic meaning.
  \item \textbf{SASRec}~\cite{kang2018self} uses uni-directional self-attention for autoregressive sequence modeling.  
  It effectively captures local dependencies but ignores item-level multimodal semantics.
  \item \textbf{FDSA}~\cite{zhang2019feature} integrates user and item attributes into the attention mechanism to enhance personalization.
  \item \textbf{S$^3$-Rec}~\cite{zhou2020s3} introduces self-supervised learning tasks to enrich representation learning,  
  yet still relies on pure ID embeddings lacking transferability.
\end{itemize}

\noindent\textbf{(3) VQ / Multimodal Models.}  
\begin{itemize}[leftmargin=*]
  \item \textbf{VQ-Rec}~\cite{hou2023learning} utilizes vector quantization for sequential recommendation.
  \item \textbf{MISSRec}~\cite{wang2023missrec} aligns text and image features using pre-extracted embeddings to improve multimodal sequential recommendation.
\end{itemize}

\noindent\textbf{(4) Prompt-driven LLM Models.}  
\begin{itemize}[leftmargin=*]
  \item \textbf{P5-CID}~\cite{hua2023index} organizes multiple recommendation tasks in a text-to-text format and models different tasks uniformly using the T5 model.
  \item \textbf{VIP5}~\cite{geng2023vip5} extends P5 into a multimodal framework using CLIP image encoders and text prompts.  
  These methods achieve strong zero-shot generalization but depend heavily on language priors instead of learned item semantics.
\end{itemize}

\noindent\textbf{(5) Semantic-ID / Generative Models.}  
\begin{itemize}[leftmargin=*]
  \item \textbf{TIGER}~\cite{rajput2023recommender} is the first to employ RQ-VAE for constructing discrete semantic item IDs, enabling generative recommendation.
  \item \textbf{MQL4GRec}~\cite{zhai2025multimodal} further quantizes multimodal embeddings into a unified ``quantitative language'' to enhance cross-domain transferability.
\end{itemize}

\section{More Ablation Study}
We futher compares our dynamic fusion strategy with a traditional fixed-layer fusion mechanism and the result is on Tabel\ref{tab:w/o_dynamic_hr5_ndcg5}
Across all three datasets, dynamic fusion consistently outperforms the traditional counterpart, yielding clear improvements in both HR@10 and NDCG@10. Specifically, the gains are most pronounced on the Arts dataset (+4.18\% HR@10 and +3.58\% NDCG@10), where item semantics exhibit higher heterogeneity. This confirms that a one-size-fits-all fusion depth is insufficient for diverse multimodal interactions.

\begin{table}[H]
\centering
\caption{Comparison between dynamic fusion and traditional fusion. Only HR@10 and NDCG@10 are reported.}
\label{tab:w/o_dynamic_hr5_ndcg5}
\resizebox{\columnwidth}{!}{
\begin{tabular}{lcccccc}
\toprule
\multirow{2}{*}{Method} 
& \multicolumn{2}{c}{Instruments} 
& \multicolumn{2}{c}{Arts} 
& \multicolumn{2}{c}{Games} \\
\cmidrule(lr){2-3} \cmidrule(lr){4-5} \cmidrule(lr){6-7}
& HR@10 & NDCG@10
& HR@10 & NDCG@10 
& HR@10 & NDCG@10 \\
\midrule
Traditional Fusion
& 0.1387 & 0.1067
& 0.1362 & 0.1003 
& 0.1043 & 0.0572 \\
Ours (Dynamic Fusion) 
& \textbf{0.1418} & \textbf{0.1077} 
& \textbf{0.1419} & \textbf{0.1039} 
& \textbf{0.1061} & \textbf{0.0575} \\
\bottomrule
\end{tabular}
}
\end{table}

\begin{table}[H]
\centering
\caption{Model complexity and inference efficiency across datasets. The table reports the total floating-point operations (FLOPs), trainable parameters, and average per-sample inference latency.}
\label{tab:model_complexity}
\begin{tabular}{lccc}
\toprule
\textbf{Dataset} & \textbf{FLOPs (M)} & \textbf{Params (M)} & \textbf{Latency (ms/sample)} \\
\midrule
Instruments & 185.36 & 3.71 & 2.66 \\
Arts        & 225.68 & 4.51 & 2.67 \\
Games       & 282.14 & 5.64 & 2.69 \\
\bottomrule
\end{tabular}
\end{table}

\section{Discussion}
We further analyze the model complexity across datasets in Table~\ref{tab:model_complexity}.
While the FLOPs and parameter counts gradually increase from Instruments (185.36M / 3.71M) to Arts (225.68M / 4.51M) and Games (282.14M / 5.64M), this growth primarily stems from larger dataset sizes and expanded vocabulary spaces, rather than architectural overhead.
Overall, our model remains lightweight (fewer than 6M parameters) and achieves efficient inference (<3 ms per sample), demonstrating strong scalability and computational efficiency for large-scale recommendation scenarios.
\end{document}